\documentclass[lettersize,journal]{IEEEtran}
\usepackage{amsmath,amsfonts}
\usepackage{algorithmic}
\usepackage{algorithm}
\usepackage{array}
\usepackage[caption=false,font=normalsize,labelfont=sf,textfont=sf]{subfig}
\usepackage{textcomp}
\usepackage{stfloats}
\usepackage{url}
\usepackage{verbatim}
\usepackage{graphicx}
\usepackage{cite}
\usepackage{lineno,hyperref}
\modulolinenumbers
\usepackage{amsmath}
\usepackage{soul}
\usepackage{color}
\usepackage[usenames,dvipsnames]{xcolor}
\usepackage{multirow}
\usepackage{longtable}
\usepackage{graphicx}
\usepackage{longtable}
\usepackage{tikz}
\usepackage{amsmath}

\usepackage{booktabs} % For prettier tables
\usepackage{longtable}
\usepackage{xurl}
\PassOptionsToPackage{hyphens}{url}\usepackage{hyperref}
\def\checkmark{\tikz\fill[scale=0.4](0,.35) -- (.25,0) -- (1,.7) -- (.25,.15) -- cycle;} 
\begin{document}
\title{EarlyMalDetect: A Novel Approach for Early Windows Malware Detection Based on Sequences of API Calls}

%\title{A Sample Article Using IEEEtran.cls\\ for IEEE Journals and Transactions}
\author{Pascal  Maniriho,
        Abdun Naser Mahmood,
        and~Mohammad Jabed Morshed Chowdhury % <-this % stops a space
%IEEE Publication Technology,~\IEEEmembership{Staff,~IEEE,}
        % <-this % stops a space
\thanks{P. Maniriho, A. N. Mahmood, and M.J.M. Chowdhury are with the Department
of Computer Science and Information Technology, La Trobe University, Melbourne,
%VIC, 3083 Australia e-mail: \{p.maniriho, a.mahmood, m.chowdhury\}@latrobe.edu.au.}% <-this % stops a space
%\thanks{Manuscript received January 20, 2024; revised January 20, 2024.
}}

\markboth{Journal of \LaTeX\ Class Files,~Vol.~14, No.~8, August~2021}%
{Maniriho \MakeLowercase{\textit{et al.}}: EarlyMalDetect: A Novel Preventive Approach for Early Malware Detection from Sequences of API Calls}


\maketitle

\begin{abstract}
In this work, we propose EarlyMalDetect, a novel approach for early Windows malware detection based on sequences of API calls. Our approach leverages generative transformer models and attention-guided deep recurrent neural networks to accurately identify and detect patterns of malicious behaviors in the early stage of malware execution. By analyzing the sequences of API calls invoked during execution, the proposed approach can classify executable files (programs) as malware or benign by predicting their behaviors based on a few shots (initial API calls) invoked during execution. EarlyMalDetect can predict and reveal what a malware program is going to perform on the target system before it occurs, which can help to stop it before executing its malicious payload and infecting the system. Specifically, EarlyMalDetect relies on a fine-tuned transformer model based on API calls which has the potential to predict the next API call functions to be used by a malware or benign executable program. Our extensive experimental evaluations show that the proposed approach is highly effective in predicting malware behaviors and can be used as a preventive measure against zero-day threats in Windows systems. 
 
\end{abstract}

\begin{IEEEkeywords}
Dynamic malware analysis, Malware detection, Transformer model,  Recurrent neural networks, API Calls, Transfer learning, Machine learning, Deep learning 
%Article submission, IEEE, IEEEtran, journal, \LaTeX, paper, template, typesetting.
\end{IEEEkeywords}

\section{Introduction}
\label{introduc}
\IEEEPARstart{T}{}\uppercase{HE} prevalence of Internet technologies provides many people with access to a diverse range of services, including social networking, online retail, navigation, and object positioning through the Global Positioning System (GPS), among others. As reported by Petrosyan \cite{Internet90:online}, there are 5.35 billion Internet users around the globe as of April 2024, which accounts for 66.2\% of the world’s population. DataReportal \cite{DigitalA1:online} also revealed that Internet users are on the rise with statistics showing that the number grew by 100 million users in 2022. These billions of users connected worldwide make the Internet the center and pillar of today's digital world.

However, it is worth mentioning that despite the benefits of increased connectivity through the Internet, there is also a higher risk of cyber attacks due to the growing number of interconnected devices and systems. Hackers develop advanced malware programs with new ways to infiltrate systems and steal sensitive information, posing a growing risk to computer security worldwide \cite{Internat7:online}. Furthermore, hackers are constantly innovating their attack mechanisms, making it increasingly difficult for traditional security techniques to keep up. As reported in \cite{Ransomwa24:online}, victims of ransomware attacks have paid out a high ransom (449 million US dollars) in the first six months of 2023. Thus, taking preventive measures to secure online devices and sensitive information becomes a security requirement. 

Detecting malware based on behavioral data gathered during execution can identify advanced malware. However, it requires a long time to gather all malware behaviors, increasing the likelihood for malware to deliver the malicious payload before detection. Furthermore, the collection of behavioral data itself is computationally prohibitive and resource expensive. Many of the current detection models mostly rely on post-benign and malware execution logs (features) to identify malware activities \cite{herrera2023dynamic} \cite{qiang2022efficient} \cite{li2022novel}. Consequently, these models can detect malware activities in the aftermath of a malware infection in the system, which is already too late for safety-critical systems. This is a major limitation observed in most behavior-based malware detection models found in the literature \cite{10299708} \cite{amer2020dynamic} \cite{suaboot2020sub} \cite{9715123} \cite{10177752}.

Detecting and classifying malware in its early stages is a critical aspect of security, particularly when it comes to critical systems and networks. For Windows malware detection, the common features extracted from executable files include API call sequences \cite{li2022novel} operation code sequences (opcodes) \cite{kakisim2022sequential}, loaded dynamic linked libraries (DLLs) \cite{yousuf2023windows}, and static images of executable file \cite{sl2019windows}. Additionally, other features such as file changes, network connections history, registry changes, printable strings, and information from file headers have been also used to distinguish benign from malware applications \cite{singh2021survey} \cite{maniriho2022study}. In particular, API call features have become popular in Windows malware detection \cite{maniriho2023systematic}.

In the context of API call-based malware detection, existing techniques lack preventive techniques for early malware detection. Thus, this work is specifically focused on addressing this challenge by developing a new malware detection technique that utilizes features of API calls. The proposed method allows for effective early detection of malicious files by predicting the next possible actions to be performed by a Windows executable program (file) based on its invoked API calls. Our approach involves analyzing API calls to predict and identify threats at an early stage which allows taking appropriate action before the malware can cause damage. We rely on a sequence prediction approach based on a fine-tuned transformer model to predict the behavior of malware while the detection technique uses the predicted behavioral data to detect malware. The key advantage of the proposed approach lies in its ability to predict the behaviors of malware or benign programs since API calls are overloaded and overused in different contexts when performing various tasks in Windows OS. By predicting and classifying suspicious API calls that are indicative of malicious intent, we can quickly and effectively detect a malware program in its early execution stages, minimizing the likelihood of potential damage caused by the infection or compromise. More specifically, in this work, we aim to address the following two main research questions \textbf{(RQs)} that have not been covered in prior works based on API call sequences in Windows.

\begin{itemize}
\item Predicting sequences of API calls \textbf {(RQ1)}: Based on the initial API calls invoked by an executable program (malware or benign)  in Windows, is it possible to predict the next sequences of API calls to be invoked by that program?

\item Detecting malware at its early stage of execution \textbf{(RQ2):} Can malware be detected by predicting its behaviors before it infects the system?

\end{itemize}

More importantly, examining both research questions enables a focused investigation into the potential utility of predictive models in identifying malware through anticipated behavioral patterns of API calls preceding actual infection.

\subsection*{The Main Contributions of This Work}
The following are key contributions of this article, with each contribution underscoring important advancements and perspectives in the evolving landscape of malware detection. 

%\begin{enumerate}[leftmargin=*]
\begin{enumerate}
 
\item We design and build a transformer-based model for predicting the next API calls invoked by an executable program by fine-tuning the existing Pre-trained Transformer Model (GPT-2) using the Huggingface open-source Transformers library. The fine-tuned model (Winapicallseq-finetuneGPT2-Model) can predict the behaviors of an executable program based on initial sequences of API calls.
    
\item We propose a new malware detection approach that relies on the fine-tuned transformer model and bidirectional recurrent neural networks with an attention mechanism. The proposed detection approach can detect and classify malicious activities before they occur, making it a preventive and mitigation approach for early malware detection in Windows systems.
    
\item We perform various experiments using various datasets of API call sequences and the experimental results demonstrate valuable performance achieved by the proposed approach under different experimental conditions. The overall performance shows promising results, and the proposed approach outdid other state-of-the-art malware detection approaches tested on the same datasets.  
   
\end{enumerate}

\textbf{Organization:} The remaining part of this paper is organized as follows. Section \ref{backg} presents the background, Section \ref{related-works} discusses the related works on malware detection while Section \ref{proposed-aproach} presents the proposed approach. Section \ref{experimental-results} discusses the experimental results and limitations of the proposed approach and discusses future work. Furthermore, the conclusion of this work is presented in Section \ref{concl-future}. 

\begin{figure}[t!]
  \centering
  \includegraphics[width=0.90\linewidth]{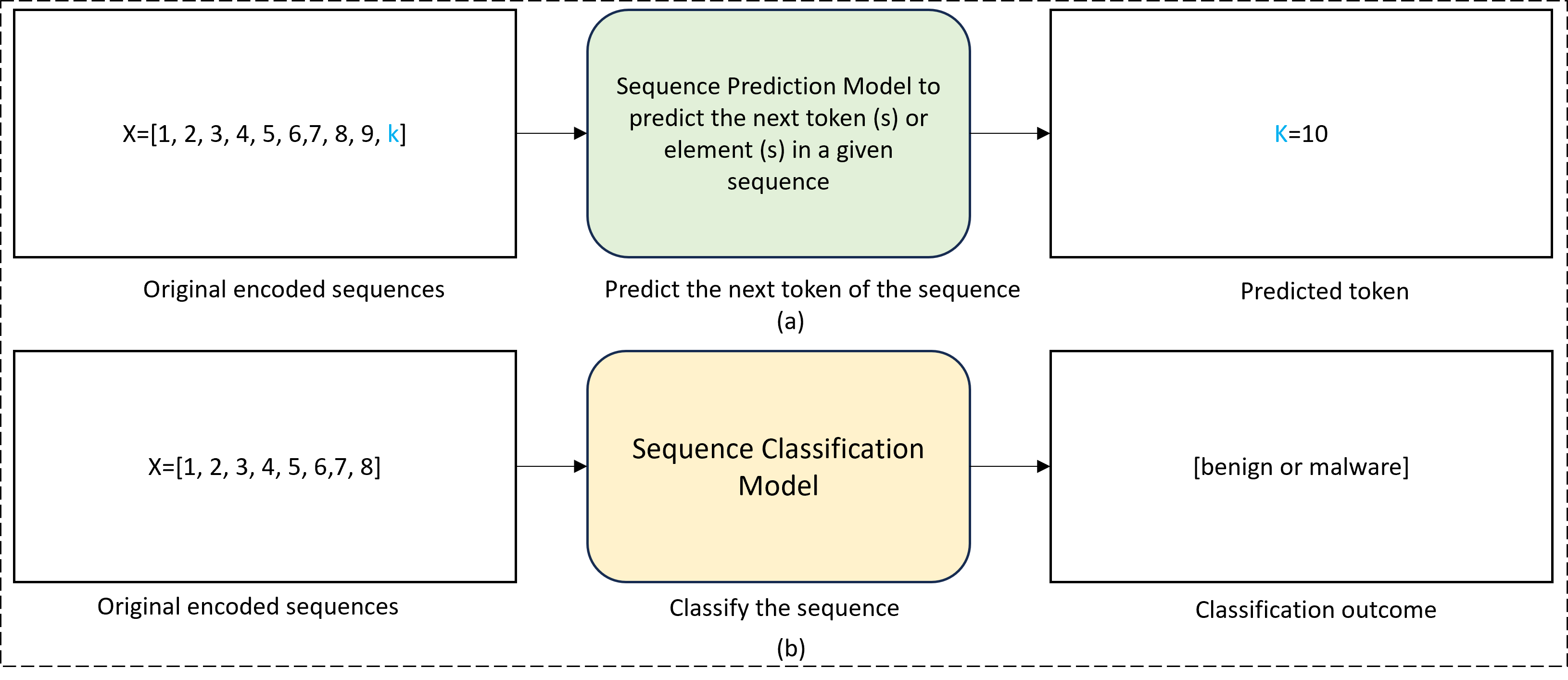}
  \caption{Modeling sequences (a) Sequence prediction  (b) Sequence classification.}
  \label{sequencemodel}
\end{figure}

\section{Background}
\label{backg}
In this section, we introduce the concepts of sequence prediction, sequence classification, and transformer models, as these techniques are foundational for building the proposed approach for early malware detection. 

\subsection{Sequence Predictions}
\label{seq-pred}
The attempt to predict elements (or words) of a sequence on the basis of the preceding elements is referred to as sequence prediction in natural language processing (NLP) \cite{sun2001sequence}. Given a sequence $X$ of text represented in the form of a sentence, the goal of a sequence prediction model is to predict the next word in the sequence based on the elements of the original sequence $X$. For instance, if the sequence $X=[1,2,3,4,5,6,7,9,k]$ with numbers indicating the encoded words from $X$, the goal is to predict the next word $k$ using a sequence prediction model $P$. It is worth mentioning that in many cases, more than one word can be predicted at the same time. A typical sequence prediction model is first trained using a deep learning algorithm and a set of sequences in the train set. Once the model has been trained, it can be leveraged to make sequence predictions. Deep learning algorithms have shown promising results in sequence predictions and have been successfully employed to build powerful language modeling models \cite{otter2020survey} \cite{lauriola2022introduction}. Given its potential, sequence prediction has gained popularity in several application domains such as speech recognition, text classification, stock market prediction, product recommender systems, web page prefetching, and many more. More importantly, Figure \ref{sequencemodel} (a) presents a typical scenario for sequence prediction.

\subsection{Sequence Classification}
\label{seq-class}
The primary goal of sequence classification is to create a classification model using a labeled dataset \cite{tang2014data}. It involves applying different feature extraction techniques to extract meaningful features from input sequences and use them to create classification models that accurately categorize and classify sequences in their respective categories. Thus, a sequence classification model aims to predict a class (label) for a given input sequence. For instance, having an encoded sequence $X=[1,2, 3,4,5,6,7]$ representing behaviors (e.g. sequence of API calls) extracted from an executable program in Windows, the task is to predict if $X$ represents a malicious file or benign file (in the context of malware classification). Accordingly, Figure \ref{sequencemodel} (b) illustrates the process for sequence clarification.

\begin{table*}[ht!]
\centering
\caption{Recent API call sequence-based approaches for detecting malware in Windows.}
\label{related-works-trends}
\scalebox{0.87}{
 \begin{tabular}{p{13mm}p{37mm}p{25mm}p{47mm}p{47mm}}
%\begin{tabular}{|l|l|l|l|l|l|}
\hline
Reference & ML/DL Algorithm& Extracted Features & Performs Automatic API Call Prediction & Performs Early Malware Detection \\ \hline
\cite{li2022novel} & CNN-BiLSTM & API call sequences  & x &  x \\ \hline
\cite{qiang2022efficient} & CNN and LSTM  & API call sequences  & x & x \\ \hline
\cite{10206378} & TextGCN & API call sequences  & x &  x  \\ \hline
\cite{kishore2023efficient} & CNN  and RF & API call sequences  & x &  x \\ \hline
\cite{lv2023ctimd} & TextCNN & API call sequences  & x & x \\ \hline
\cite{daeef2023features} & RF, SVM, GCN, and LSTM & API call sequences  & x & x  \\ \hline
\cite{zhang2023dynamic} & TextRNN & API call sequences  & x  &  x \\ \hline
\cite{avci2023analyzing} & Stacked LSTM & API call sequences  & x &  x \\ \hline
\cite{aboaoja2023dynamic} & XGBoost, RF, SVM and ANN & API call sequences  & x &  x \\ \hline
\cite{10309174} &BiLSTM, BiGRU, etc.,& API call sequences& x &  x \\ \hline
\cite{10063366} & BERT-GCN & API call sequences  & x  & x  \\ \hline
This work &DistillBert-BiGRU-Attention& API call sequences  & \checkmark  & \checkmark  \\ \hline
\end{tabular}
}
\end{table*}

%%%%\cite{9936578} & SVM and LR & \API call sequences & x &  x \\ \hline

\section{Related Work}
\label{related-works}
%Behavior-based malware detection in Windows often relies on behavioral features extracted from executable files while running.

%Several approaches were proposed for detecting malware using behavioral (dynamic) features in recent years \cite{USMAN2021124} \cite{maniriho2021study}. These approaches utilize specific features extracted from executable files to discover malware.

%%%%Among these features, API Calls have been widely used in the literature due to their potential to distinguish malware from benign files. Accordingly, the in-depth systematic survey carried out in \cite{maniriho2023systematic} shows that API Calls are the most used behavioral features for malware detection in Windows OS. Thus, in this section, we discuss various approaches for detecting malware based on API calls.

Several approaches were proposed for detecting malware using behavioral (dynamic) features in recent years \cite{USMAN2021124} \cite{maniriho2021study}. These approaches utilize specific features extracted from executable files to discover malware. XiaoFeng et al. \cite{xiaofeng2019assca} combined machine learning (ML) and deep learning (DL) algorithms to implement an API call-based technique for malware detection. Their technique is based on analyzing dependency relations in sequences of API calls and relies on bidirectional residual neural networks to detect malicious programs. Using this technique, they were able to achieve a detection accuracy of 96.7\%. Li et al. \cite{li2022dmalnet} proposed DMalNet, a graph-deep neural-based framework that uses a hybrid feature encoder and call graph to extract semantic features from API call sequences and their arguments to detect Windows malware. The evaluations conducted using over 20,000 benign and 18,000 malicious executable files show a detection accuracy of 98.43\%. Ndibanje et al. \cite{ndibanje2019cross} have used statistical methods to analyze the frequency and similarity between API calls extracted from malware and benign executable files. While their similarity matching-based method achieved an accuracy of 96.7\%, the effectiveness of this approach is limited as adversaries can manipulate the frequency of API call tokens by adding superfluous (unnecessary) API calls to avoid detection. Karbad and Debbabi \cite{karbab2019maldy} employed the bag-of-words (BoW) method to transform API call sequences into word sequences and have introduced the MalDy framework for behavior-based malware detection. Although the experimental evaluations demonstrate better performance achieved with their framework, the approach used for analyzing API calls can be vulnerable to obfuscated malware which can greatly affect the effectiveness of their detection model \cite{dabas2023malanalyser}.   

Han et al. \cite{han2019maldae} proposed MalDAE, a malware detection framework that uses both static and dynamic analysis sequences of API calls. Ki et al. \cite{ki2015novel} suggested a detection technique that relies on a DNA sequence alignment approach to identify recurring patterns of  API calls across various types of malware. By examining whether specific API call functions are present on new malware samples, their approach can identify malware threats. However, it can be computationally expensive when dealing with longer sequences.    

Dabas and Prabha \cite{dabas2023malanalyser} developed MalAnalyser, a new API calls-based system for Windows malware detection. MalAnalyser works by extracting frequently occurring API calls from the entire sequences of API calls captured during execution. Dabas and Prabha’s approach has also used Particle Swarm Optimization and Genetic algorithms to identify a small subset of relevant features that enhanced the detection of unseen malware based on patterns of API calls. Amer and Zelinka \cite{amer2020dynamic} relied on behavioral graph and word embedding to analyze the contextual relationship between functions of API calls in malware detection in order to generate features fed to a Markov Chain-based approach that classifies malware. Li et. \cite{li2022intelligent} extracted API call sequences from benign and malicious files and applied a Markov chain model to extract feature vectors from  API calls by constructing a cyclic graph. The extracted features were fed to a graph convolution neural network model which achieved an accuracy rate of 98.32\% when detecting unknown malware samples. Huang et al. \cite{huang2021method} used visualization and convolutional neural networks (CNNs) to transform dynamic and static sequences of API calls into images. While this approach achieved an accuracy of 94.7\%, converting API calls directly into images or other forms of representations may lead to the loss of relevant information as noted by Ma et al. \cite{ma2019api}.

Moreover, Table \ref{related-works-trends} presents other current API call-based approaches and highlights ML or DL algorithms that were used to implement the detection model. It also highlights two main limitations of existing malware detection approaches. Accordingly, it shows if the detection approach has attempted to address the issue of automatic API call sequence prediction (predicting the behaviors of a Windows executable program when it starts executing). The last column also reveals if a given detection technique can perform early malware detection or not. In general, existing detection techniques are most likely to detect malware programs when they have already infected the system, resulting in high rates of failure. As a solution to the mentioned limitations, this work presents a new approach that performs early detection and classification of malware in Windows. 

\begin{comment}
    \begin{figure*}[ht!]
  \centering
  \includegraphics[width=0.91\linewidth]{overview-proposed_method.png}
  \caption{Overview of the architecture of the proposed approach.}
  \label{proposed-tech-methode}
\end{figure*}
\end{comment}

\section{The Proposed Malware Detection Approach}
\label{proposed-aproach}
In this work, we take advantage of sequence modeling approaches presented in Section \ref{backg} to implement a new approach (EarlyMalDetect) for early malware detection in Windows. Moreover, we also use transfer learning and attention-guided recurrent neural networks. Specifically, Figure \ref{early-maldetect} shows the architecture of the proposed methodology. In contrast to the previous malware detection approaches, EarlyMalDetect can predict the next API calls to be invoked by a malware or benign program and use them to distinguish malicious or normal activities., making it a preventive and mitigation approach with the potential to detect and stop malware programs before they can infect the target system. In the next sections, we provide more details on our methodology such as datasets used, fine-tuning procedures (steps for designing the fine-tuned model for modeling API calls), and finally the design of the detection model.

\subsection{Collecting Datasets of API Call Sequences}
\label{dataset-used}
%In Windows OS, the application programming interface (API) call sequences refer to a series of functions that an application uses to interact with the OS to perform some actions (tasks). These calls allow Windows applications to access and use various resources such as network connection services, files, hardware devices, and other systems resources such as main memory and CPU. API calls are typically initiated by an application (program) through a functional call to its code. During the execution of a program, the OS responds to these API call requests by executing the corresponding (associated) functions and returning results to the program. For example, if an application needs to create a new file, it would initiate a sequence of API calls to request permission from the OS to create the file, specify the file name and its location, and then write data to it. Such API calls may include functions like CreateFile, WriteFile, and CloseHandle. In general, API calls constitute an essential part of applications to communicate and interact with the Windows OS. This gives applications the potential to perform a wide range of operations and tasks. An example of an API call sequence extracted from a malware program (Virus) while running on Windows is presented in Figure \ref{apisequence}. Given the potential of API calls in detecting malware, 

We have collected various datasets of API calls representing malware and benign executable files. These include the API calls dataset presented in \cite{ceschin2018need}, \cite{genccaydin2022benchmark}, \cite{ki2015novel} \cite{maniriho2023api} and \cite{PDFWindo91:online}. These datasets are used for experimental evaluations (train and test the proposed approach) and details on the number of samples on each dataset are presented in Table \ref{selected-dataset}. It is worth noting that API calls are essential for applications to communicate and interact with the Windows OS, making them valuable features when building malware detection techniques in Windows.

\begin{comment}
 \begin{figure}[t!]
  \centering
  \label{apisequence}
  \includegraphics[width=0.98\linewidth]{Apicalls-sequence1.png}
  \caption{An example of a sequence of API calls extracted from a malicious executable file while executing in Windows.}
\end{figure}
\end{comment}

\begin{table}[t!]
\centering
\caption{Datasets of API call sequences collected for training and testing the proposed malware detection approach.}
\label{selected-dataset}
\scalebox{0.84}{
 %\begin{tabular}{p{5mm}{17mm}p{30mm}p{22mm}}
\begin{tabular}{llll}
\hline   
\# & Dataset Taken From & Number of Samples & File Format \\ \hline
1  & \cite{ki2015novel} &  23,438 & .Exe file   \\ \hline

2  & \cite{ceschin2018need} & 21,105 & .Exe file   \\ \hline

3  & \cite{genccaydin2022benchmark} & 24,435 & .Exe file   \\ \hline

4  & \cite{maniriho2023api}   &  2,570   & .Exe file   \\ \hline 

5   & \cite{PDFWindo91:online}  &  551  & .Exe file   \\ \hline

\end{tabular}
}
\end{table}

\subsection{Fine-tuning the GPT-2 Transformer Model on API Calls Dataset Through Transfer Learning}
\label{finetune-model}
 Training deep neural networks like transformer models \cite{10.5555/3295222.3295349} from scratch can be challenging due to several factors. For instance, obtaining the necessary amount of data for the problem at hand can be time-consuming. Additionally, acquiring the computational resources (such as GPUs) required to train these deep learning models can also be costly \cite{AnIntrod66:online} \cite{yang2020transfer}. As a result, one way to overcome the challenge is by utilizing the transfer learning approach which offers various benefits such as reducing training time, accelerating the training process for new models, and minimizing the model deployment time. Transfer learning allows us to use a pre-trained model (a model trained on large-scale data from a source domain/non-specific domain) as a starting point for building a new model \cite{yang2020transfer}. 

Although transfer learning has gained more popularity in fields such as computer vision over the past years, it has not been fully explored in other fields such as cybersecurity, particularly in malware detection. Some of the previous works attempted to use transfer learning to build malware detection approaches based on images of Windows executable files. For example, Kumar and Janet \cite{kumar2022dtmic} achieved a good detection accuracy (98.92\%) by using various CNN-based pre-trained models (like Google’s inceptionV3, VGG16, ResNet50, and VGG19) as feature extractors. Lo et al. \cite{8763852} examined the performance of Xception and  VGG16 models on both Malimg and Microsoft Malware image Datasets. Nevertheless, existing techniques did not explore the use of transfer learning in API call-based malware detection, specifically in the automatic prediction of API call sequences. 

\begin{figure}[!]
  \centering
  \label{finetune-gpt-appicalls}
  \includegraphics[width=0.90\linewidth]{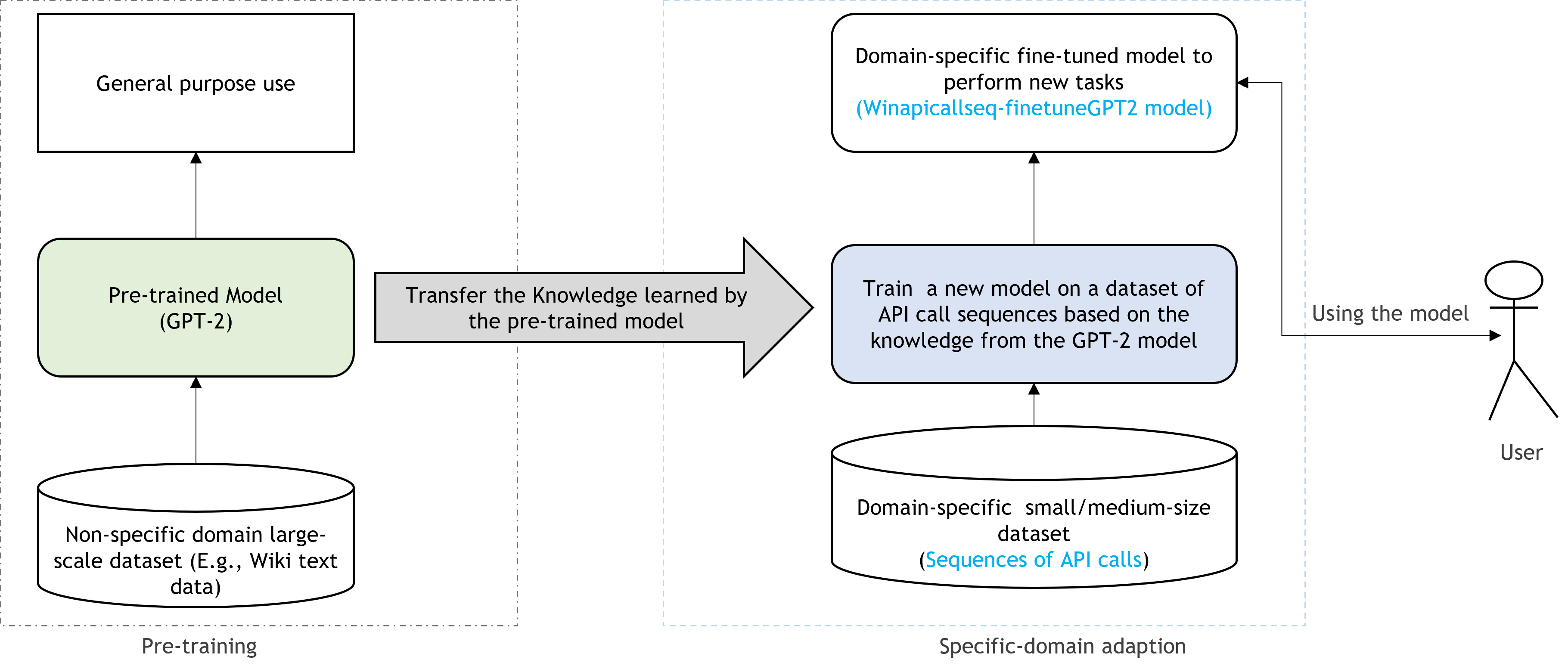}
  \caption{The process for fine-tuning the GPT-2 model on a dataset of API call sequences through transfer learning.}
\end{figure}

Thus, we exploit the potential of transfer learning in this work. Specifically, we introduce Winapicallseq-finetuneGPT2-Model, a new domain-specific transformer model for modeling API call sequences. Our model is built by fine-tuning the GPT-2 transformer model on our custom dataset of API calls representing malware and benign executable files. GPT-2 is an open-source generative pre-trained transformer model introduced by OpenAI in \cite{radford2019language}. It was implemented and made publicly accessible by the Hugging Face Community \cite{HuggingF79:online}. GPT-2 is a powerful language model that has been pre-trained on a large corpus of text and it can perform new text generation, text data augmentation, and next-word or sentence predictions in a format similar to the content (text) it was trained on. However, it is not an optimal choice (most appropriate option) for performing specific tasks such as modeling the sequence of API calls. This is because the corpus of text (natural language text) used to train the original pre-trained GPT-2 model is different from the executable program's API call sequences. That is, API call sequences (also called the sequence of function calls) are not typically expressed in natural language. Even if some of the API call function names may use words that are part of the natural language, the overall structure of sequences of API calls is not expressed in a typical natural language sentence format or structure. 

\begin{algorithm} [ht!]
\caption{: Fine-tuning GPT-2 on API calls dataset} \label{alg1}
\begin{algorithmic}
\STATE \textbf{Notations:} 
\STATE $P:$ Required package 
\STATE $F:$ Input file with custom dataset 
\STATE Output: Fine-tuned model 
\STATE \textbf{Perform these steps}
\STATE 1. Load all required $P$ 
\STATE 2. Load the GPT-2 model
\STATE 3. Load the GPT-2 tokenizer
\STATE 4. Define data processing function
\STATE 5. Load and pre-process $F$
\STATE 6. Split the data into training and test sets
\STATE 7. Define hyperparameter for training
\STATE 8. Train the GPT-2 model on processed F
\STATE 9. Evaluate the trained model
\STATE 10. Save the fine-tuned model 
\STATE 11. Terminate the script
\end{algorithmic}
\end{algorithm}

Consequently, we have fine-tuned the GPT-2 model on the datasets of API call sequences to produce Winapicallseq-finetuneGPT2-Model, a domain-specific model for modeling sequences of API calls in Windows. The model was fine-tuned using 68,961 sequences of API calls from the dataset presented in \cite{ki2015novel} \cite{ceschin2018need} and \cite{genccaydin2022benchmark}. The fine-tuning process was performed following the guidelines outlined in \cite{Finetune25:online} using the Hugging Face Transformers library and PyTorch framework. This involves processing our dataset of API call sequences by converting and assembling them into input batches of tensors supported by the GPT-2 transformer model using a tokenizer from the transformers library. The tokenizer splits the sequences of API calls into tokens. These tokens of API calls also get converted into numerical representation and then tensors which serve as the input for the model. It is worth noting that we have also applied sequence padding and truncation methods to handle variable lengths of API calls. This is important as the model requires all tensors to have the same length.

Training is the next stage after data pre-processing. Thus, we have used the trainer class optimized for training transformers models which is provided in the transformers library \cite{HuggingF79:online}. This makes training easier and avoids manually writing a training loop from scratch. We trained the model with default parameters as outlined in the training configuration class presented in \cite{OpenAIGP78:online}. However, we have set the maximum block size/sequence length to 500. The fine-tuned model is used to predict the next API calls to be invoked by a Windows executable program (malware or benign) during execution. In addition, the fine-tuned model can also be used to generate new sequences of API calls. We make the fine-tuned model available for the research community and it can be freely accessed from \cite{mpascoEa88:online}. It is also important to mention that the labels of the samples were discarded during training, allowing us to fine-tune the model through unsupervised learning. This is because our goal is not to use the fine-tuned model for performing any classification tasks. Thus, labeled API call sequences are only considered when building the detection model, as discussed in subsection \ref{desin-detect}. More details on the fine-tuning steps are presented in Figure \ref{finetune-gpt-appicalls}, while the main steps can also be found in the algorithm presented in \ref{alg1}. In addition, Section \ref{earlymaldetect-withproposed} discusses how the fine-tuned transformer model is used to discover the behaviors of an executable file by predicting its next API calls to be used while executing.  

\begin{figure*}[ht!]
  \centering
  \label{early-maldetect}
  \includegraphics[width=0.92\linewidth]{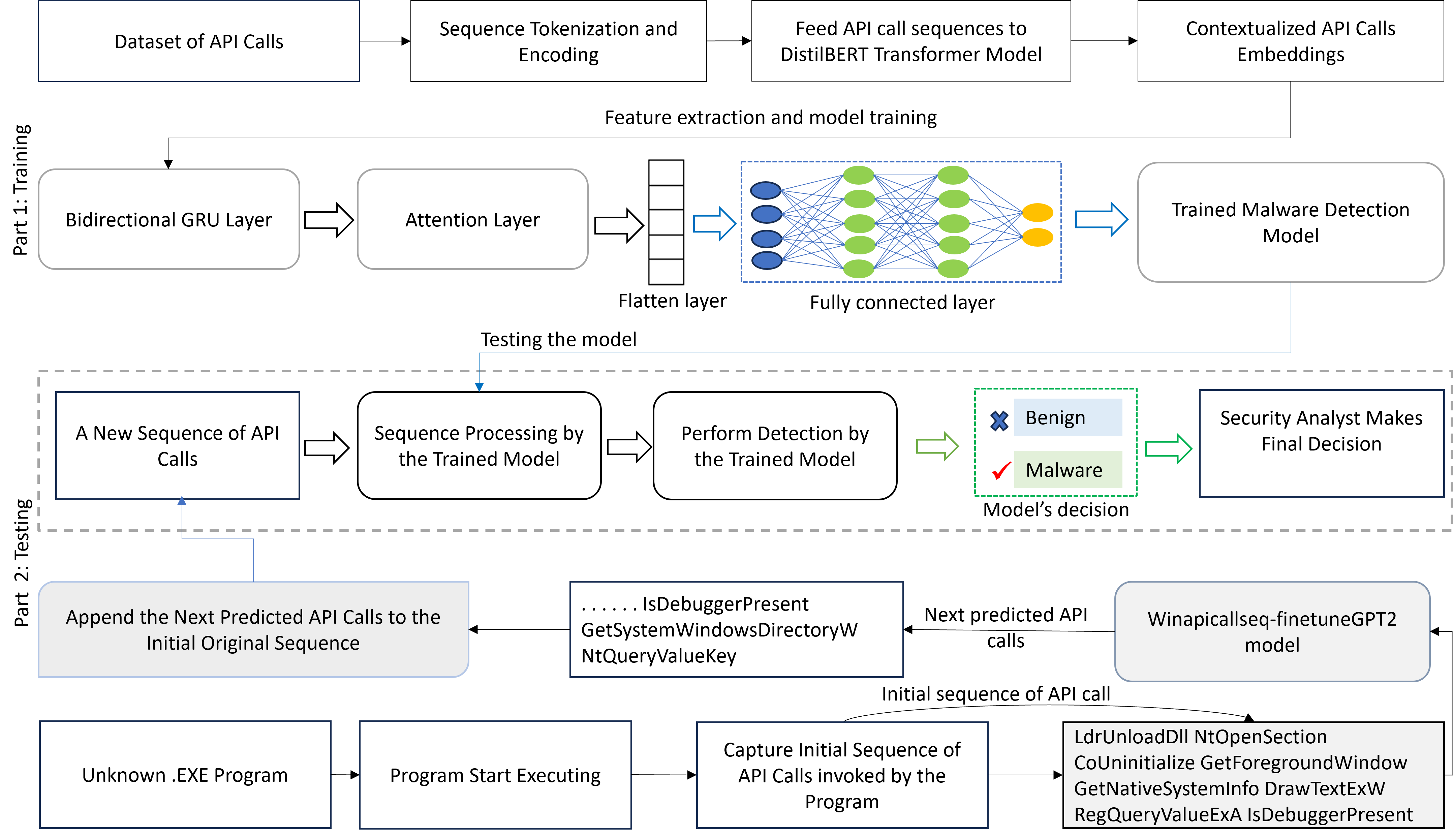}
  \caption{The proposed approach for early malware detection based on the fine-tuned transformer model.}   
\end{figure*}

\subsection{Designing the Detection Model}
\label{desin-detect}
The proposed detection model is designed during this phase and it involves training and testing as illustrated in Figure \ref{early-maldetect} part 1 and part 2. Training is performed using the datasets of API call sequences from \cite{ki2015novel} \cite{ceschin2018need} while testing is carried out using the datasets in \cite{PDFWindo91:online} and \cite{maniriho2023api}. In our experiment, we refer to the dataset in \cite{PDFWindo91:online} and \cite{maniriho2023api} as dataset 1 and dataset 2, respectively. During testing, we rely on the next sequence prediction method using the fine-tuned model (Winapicallseq-finetuneGPT2-Model) in order to identify and detect unknown Windows malicious files as early as possible before they can infect and compromise the system. The testing process is illustrated in part 2 of Figure \ref{early-maldetect}, which shows the necessary steps to detect malware based on the fine-tuned model and the designed detection model. In addition, more details on testing are presented in Section \ref{earlymaldetect-withproposed}.

It is also important to mention that we have considered the best practices for training and testing malware detection models suggested in the works in \cite{pendlebury2019tesseract} \cite{maniriho2023api}, allowing us to follow practical procedures to avoid temporal and spatial bias in our experiments. The proposed detection model is also based on the DistilBERT transformer model which encodes and embeds sequences of API calls. That is, DistilBERT creates numerical representations of contextual API call embeddings (also known as embedding matrix) which are passed to a bidirectional gated recurrent unit with attention (BiGRU-Attention) that learns and extracts contextual patterns of API calls. The output from the BiGRU-Attention layer is then passed to the last layer with fully connected neurons (fully connected layer) that performs further non-linear processing followed by classification of the sequence. The output of the fully connected layer reveals if the API call sequence represents malware or a legitimate Windows application. More specifically, sections \ref{process-api}, \ref{bigur} \ref{attention-layer} and \ref{dnnslayer} provide details on each part of the proposed malware detection approach and discuss how they interact to perform early malware detection. 

\subsubsection{API call Sequence Numerical Representation}
\label{process-api}
Before feeding sequences of API calls to the feature extraction layer (BiGRU-Attention), we have to represent tokens of API calls as numbers. We refer to this process as generating numerical representation and it is accomplished through API calls tokenization and embedding. Tokenization involves breaking down a sequence of API calls into individual tokens of API calls. During tokenization, each unique API call in the sequence is given a numerical index, with the most common API call tokens having lower indices. This is typically done to create a fixed-size vocabulary of the most frequent API calls in the corpus. Because API call sequences in the dataset have different lengths, they have to be standardized to allow them to have the same length. Thus, in this work, this is achieved by padding shorter API call sequences with zeroes and truncating longer sequences to a maximum length that is defined. After tokenizing, indexing, and padding input sequences of API calls, each token of an API call is then represented as an embedding vector through embedding.

In this work, we employ the DistilBERT model \cite{sanh2019distilbert} to perform tokenization, padding, and truncation of API call sequences and to create their embeddings. DistilBERT is a smaller and faster variation of the BERT transformer architecture that utilizes distillation to shrink the larger BERT model (reduce its size and complexity). In comparison to the BERT model, DistilBERT has 40\% few parameters while maintaining over 95\% of BERT's performance on the GLUE language understanding benchmark. Additionally, this smaller lightweight model runs 69\% faster than its predecessor, making it preferable in our experiment. Consequently, we have trained the DistilBERT on our API calls dataset to create contextualized API call embedding features that improve the detection of malware attacks. During tokenization and padding, DistilBERT generates an attention mask to indicate which tokens are actual API calls versus padding. All padded sequences are processed by DistilBERT to generate embeddings for API calls. Additionally, the parameter configurations for the  DistilBERT model are provided in Table \ref{distil-config}. Once all API calls are encoded, the embedding vectors are then passed to the next layer for contextual feature extraction. 

\begin{flalign}
\label{fcn-relu}
&l_{i}=ReLU(\sum_{i}^{}W_{i}*h{_{t}{i}}+b_{i})&
\end{flalign}

\begin{flalign}
\label{sig}
&Sigmoid (x) =\frac{1}{1+e^{-x}}&
\end{flalign}

\begin{table}[ht!]
\centering
\caption{DistilBert model configurations parameters}
\label{distil-config}
%\resizebox{\textwidth}{!}{%
%\begin{tabular}{ll}
\scalebox{0.91}{
 \begin{tabular}{p{45mm}p{37mm}}
\hline
Parameter name            & Value                 \\ \hline
Activation function       & Gelu                  \\ \hline
Architecture              & DistilBertForMaskedLM \\ \hline
Attention\_dropout        & 0.1                   \\ \hline
Embedding dimension       & 768                   \\ \hline
Dropout                   & 0.1                   \\ \hline
Hidden \_dim              & 3072                  \\ \hline
Initializer\_range        & 0.02                  \\ \hline
Max\_position\_embeddings & 512                   \\ \hline
Model\_type               & Distilbert            \\ \hline
N\_heads                  & 12                    \\ \hline
n\_layers                 & 6                     \\ \hline
Pad\_token\_id            & 0                     \\ \hline
Sinusoidal\_pos\_embds    & False                 \\ \hline
Tie\_weights              & true                  \\ \hline
Vocab\_sizE               & 30522                 \\ \hline
\end{tabular}%
}
\end{table}

\subsubsection{Bidirectional GRU Layer}
\label{bigur}
In the proposed detecting model depicted in Figure \ref{early-maldetect} part 1, the bidirectional gated recurrent unit (BiGRU) layer takes the contextual API calls embeddings matrix generated by the DistilBERT transformer model as input. This layer then processes the input sequence in both forward and backward directions to capture more context from both directions of the input sequence. By using forward and backward sequence modeling, this layer learns and captures more dependencies between API calls in a given sequence (with the sequence representing a legitimate or malware program) and improves the ability of the detection layer (fully connected layer in our case). This allows the detection model to make accurate predictions when classifying malware attacks. Specifically, the BiGRU layer is configured with 32 hidden units which control the dimensionality of the output space. The network is activated using the Sigmoid and ReLU activation functions while a Dropout regularizer with a rate of 0.3 is used to prevent model overfitting. The Sigmoid and ReLU activation are computed using the equations in \eqref{fcn-relu} and \eqref{sig}. The output of bidirectional GRU contains hidden states representing API calls which are fed to the next layer of the proposed early malware detection approach.   

\subsubsection{Attention Layer}
\label{attention-layer}
The attention layer is used in our detection model's architecture to weigh the input tokens before passing them to the last layer. Specifically, the attention layer receives the input processed by the BiGRU layer and then computes the importance of each input token. This involves computing the attention weights for each input token by performing a dot product between the concatenated output from the bidirectional BiGRU layer. The attention layer uses a self-attention mechanism, where the weights are solely calculated based on the input tokens themselves. This process helps the proposed model to focus on relevant parts of the input sequences of API calls, allowing the model to produce more accurate predictions. On the other hand, the flatten layer in Figure \ref{early-maldetect} is used to convert the 3-dimensional output tensor of the attention layer into a 2-dimensional tensor suitable for the Softwamx activation function. Thus, the attention layer is followed by a Softmax activation function which is applied to the flattened tensors to obtain the final attention weights.

%%%% The Softmax is used instead of Sigmoid as the output of the attention layer is a sequence of weights corresponding to each token in the input sequence of API calls. In this case, the use of Softmax activation is to normalize these attention weights so that they add up to 1 (ensure their sum is equal to 1), allowing the network to allocate attention appropriately based on the relative importance of each token. That is, this ensures that the attention weights represent a valid probability distribution over the input tokens in a given sequence of API calls. 

\subsubsection{Fully Connected Layer}
\label{dnnslayer}
The output tensors from the attention layer are passed to a fully connected layer of neural networks which learns from them to perform the classification of malware. During training, each input sequence passes through hidden layers on the network. The process involves adjusting the weights and biases between neurons of the network through backpropagation, to optimize its parameters and minimize the learning loss which is measured using Binary Cross-Entropy presented in equation \eqref{log-loss}. Our fully connected layer's architecture is designed with two hidden layers and one output layer. The first hidden layer has 64 neurons with ReLU activation function and takes the API call context vector from the attention layer as input. The second layer has 32 neurons with the ReLU activation function while the output layer has one neuron with the Sigmoid activation function. The output ranges between 0 and 1, indicating if the input sequence belongs to the negative class (benign) or positive class (malware), respectively. After this stage, the output is a trained malware detection model that can be tested on unseen samples to make predictions. Thus, in the next section \ref{earlymaldetect-withproposed} we discuss how the trained model works in combination with the fine-tuned model presented in section \ref{finetune-model} to perform early malware detection as illustrated in Figure \ref{early-maldetect}.

\begin{figure*}[!]
  \centering
  \label{apicalls-pred-gen}
  \includegraphics[width=0.92\linewidth]{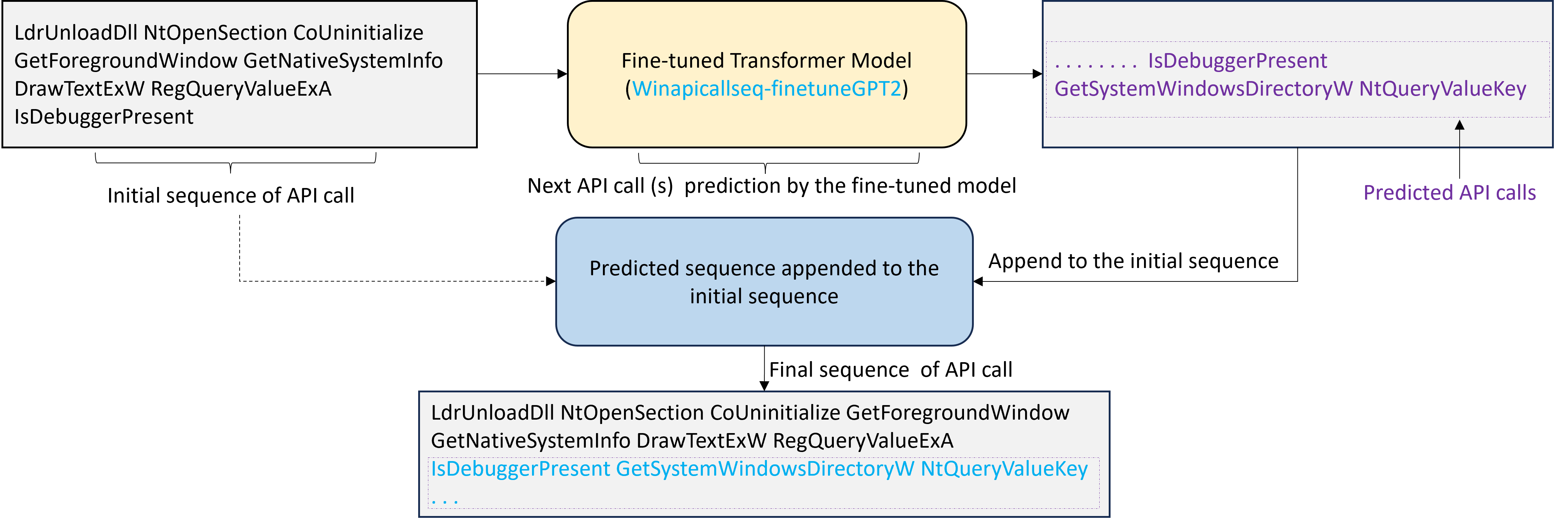}
  \caption{The process for predicting the next API calls given the initial sequence invoked by an executable file during execution.}
\end{figure*}

\begin{algorithm} [ht!]
\caption{: Predicting the next API calls} \label{alg2}
\begin{algorithmic} 
\STATE \textbf{Notation:}
\STATE $S:$ Initial sequences of API calls
\STATE $Sn:$ Next API calls to be predicted
\STATE $P:$ Required library or packages
\STATE $S': S$ Combined with predicted API calls
\STATE $F:$ Input file containing $S$
\STATE $F':$ Output file containing $S'$
\STATE $M:$ Fine-tuned model
\STATE \textbf{Function:}
\STATE Load all required $P$ 
\STATE Load $M$
\FORALL {S in F}
   \STATE Feed each $S$ to $M$
   \STATE Predict $Kn$ using $M$
   \STATE Append $Kn$ to $S$ to get $S'$
   \STATE $S' \leftarrow {Kn+S}$ 
   \STATE Save each $S'$ to $F'$
\ENDFOR
\end{algorithmic}
\end{algorithm}

\subsection{Testing the Trained Model}
\label{earlymaldetect-withproposed}
As shown in Figure \ref{early-maldetect}, the second part of our methodology deals with testing the proposed approach to perform early malware detection based on sequence prediction and classification. More specifically, we fully rely on our fine-tuned Winapicallseq-finetuneGPT2-Model model to perform early malware detection. Having an initial short sequence of API calls made by a malware or benign program when it starts executing in Windows, we use the fine-tuned transformer model to predict the behaviors of that particular program. That is, the initial sequence is fed to the fine-tuned transformer model which predicts the next possible API call functions to be used by the program under investigation, allowing us to discover the malicious behaviors of malware before it can infect the system. The predicted API calls are then appended to the original (initial) sequence to create a new sequence of API calls which is then fed to our trained malware detection model to determine if the program is malicious or legitimate. Because the behaviors of a malware program can be predicted before it can execute all of its malicious payloads and perform malicious activities, there is a high probability for the proposed approach to stop the malware and prevent it from harming the victim's system. Accordingly, Figure \ref{apicalls-pred-gen} illustrates the process of  API call prediction (for predicting the behaviors of a program through API calls) using the fine-tuned model While all steps for performing API call prediction are summarized in the algorithm presented in \ref{alg2}. 

\begin{comment}
\begin{figure*}[!]
  \centering
  \label{apicalls-pred-gen}
  \includegraphics[width=0.98\linewidth]{del.png}
  \caption{Steps for generating new sequences of API calls using the proposed fine-tuned model.}
\end{figure*}  
\end{comment}

\begin{flalign}
\label{log-loss}
&LF(y,y')=-(y*log(y')+(1-y)*log(1-y'))&
\end{flalign}

\section{Experimental Evaluations and Results}
\label{experimental-results}
This section gives details on the programming environment, tools, and performance metrics. It also presents experimental results and provides insights into the overall performance of the proposed approach. 

\subsection{Environment and Tools}
\label{env-tools}
As processing large-size sequential datasets with transformer models requires a computer with Graphic Processing Units (GPUs), fine-tuning the GPT-2 model on API calls was performed in Google Colaboratory (Google Colab) \cite{WelcomeT69:online} on a computer with  51GB RAM, 16GB for GPU RAM, and 166GB disk storage with a Colab Pro subscription. Given the high cost of GPU access in Google Colab, the fine-tuned model was saved to the local disk, and the remaining experiments (designing and implementing the proposed early malware detection approach) were performed on our computer with 12th Gen Intel(R) Core(TM) i7-12700H 2.30 GHz processor, 
32.0 GB RAM, 4GB GPU RAM with NVIDIA GeForce RTX 3050 Ti, and 1TB disk storage. The transformers library from the  Hugging Face hub\cite{HuggingF79:online}, Pytorch, TensorFlow framework, and Scikit-learn frameworks were used for the implementation in Python. Jupyter Notebook was also used as the main integrated development environment (IDE).

\subsection{Measuring the Performance}
\label{metrics}
We have assessed the performance by examining the false negative rate (FNR), false positive rate (FPR), true positive rate (TPR), true negative rate (TNR), accuracy, F1-score, recall, precision, and ROC curve ( AUC-ROC) score, macro average and weighted average for recall (R), precision (P), and F1-score (F1).  These metrics help us to determine whether the proposed approach performs well when detecting malware. 

\begin{table*}[]
\centering
\caption{Classification accuracy and AUC-ROC score achieved by the proposed approach on unseen samples from dataset 1.}
\label{result1}
\scalebox{0.87}{
\begin{tabular}{p{50mm}p{50mm}p{35mm}p{22mm}p{22mm}}

%\begin{tabular}{lllll}
\hline
Length of Initial Sequence &  Next Predicted API Calls & Total No.of API Calls & Accuracy (\%) & AUC-ROC Score \\ \hline
20                          & 10                                 & 30                                     &   91.14     &  0.9413  \\ \hline
20                          & 20                                 & 40                                     &   95.39     &   0.9427  \\ \hline
20                          & 30                                 & 50                                     &   95.97     &   0.9477    \\ \hline
\multicolumn{2}{l}{Average}                                      & 40                                     &   94.05     &   0.9439    \\ \hline
\end{tabular}
}
\end{table*}

\begin{table*}[!ht]
\centering
\caption{The precision, recall, and F1-score achieved by EarlyMalDetect  using API calls sequences from dataset 1.}
\label{result2}
\scalebox{0.76}{
%\begin{tabular}{|p{28mm}|p{55mm}|p{47mm}|}
\begin{tabular}{p{38mm}p{35mm}p{30mm}p{23mm}p{20mm}p{20mm}p{20mm}p{20mm}p{20mm}}
\hline
Length of Initial Sequence & Next Predicted API Calls & Total No.of API Calls & \multicolumn{2}{c}{Precision}  & \multicolumn{2}{c}{Recall}  & \multicolumn{2}{c}{F1-Score} \\ \cline{4-9} 
&  &  & \multicolumn{1}{c}{Benign} & Malware  & \multicolumn{1}{c}{Benign} & Malware & \multicolumn{1}{c}{Benign} & Malware \\ \hline

20  & 10      &     30   & \multicolumn{1}{c}{} 0.9976  &  0.5019  & \multicolumn{1}{c}{} 0.9049    &  0.9778    & \multicolumn{1}{c}{} 0.9490  & 0.6633  \\ \hline 

20   & 20    &     40    & \multicolumn{1}{c}{} 0.9932  &  0.6562 & \multicolumn{1}{c}{} 0.9521 &  0.9333  & \multicolumn{1}{c}{} 0.9722 & 0.7706 \\ \hline 

20   & 30    &     50     &\multicolumn{1}{c}{} 0.9933    &  0.7079     & \multicolumn{1}{c}{} 0.9622    &  0.9333       & \multicolumn{1}{c}{} 0.9775 &  0.8051      \\ \hline

\multicolumn{2}{l}{Average}  & 40 & \multicolumn{1}{c}{} 0.9947  &  0.6220 & \multicolumn{1}{c}{} 0.9397 & 0.9481 &  \multicolumn{1}{c}{} 0.9662 & 	0.7463 \\ \hline

\end{tabular}%
}
\end{table*}

\begin{table*}[h!]
\centering
\caption{Macro average and the weighted average for precision, recall, and F1-score achieved by EarlyMalDetect on API calls from dataset 1.}
\label{result3}
\scalebox{0.68}{
\begin{tabular}{p{38mm}p{35mm}p{30mm}p{20mm}p{18mm}p{18 mm}p{18mm}p{18mm}p{18mm}}
\hline
Length of Initial Sequence & Next Predicted API Calls & Total No.of API Calls & Macro P & Macro R & Macro F1 & Weighted P & Weighted R & Weighted  F1 \\ \hline
20 & 10  &  30  &  0.7497  &  0.9413   & 0.8061  & 0.9533 &    0.9114  &   0.9235      \\ \hline

20 & 20 &   40   & 0.8247 &   0.9427   & 0.8714    &  0.9631  &  0.9504  & 0.9542    \\ \hline

20 & 30 &  50    & 0.8506  &  0.9478 &   0.8913 &  0.9678  &  0.9597  &  0.9621     \\ \hline

\multicolumn{2}{l}{Average} &  40 &  0.8083 & 0.9439& 0.8563 &	0.9614 & 0.9405 & 0.9466  \\ \hline
\end{tabular}%
}
\end{table*}

%%%%%%We also present a comparative performance of the proposed approach against the existing state-of-the-art approaches presented in the literature. The performance of each approach is examined using the metrics presented in subsection \ref{metrics}. 

\subsection{Results and Discussion}
\label{resutl-discuss}
This section presents and discusses the results of our experiments. We conducted extensive evaluations and have carefully analyzed and interpreted the results. As described earlier, our approach relies on the fine-tuned Winapicallseq-finetuneGPT2 model which works in combination with DistilBERT, Bi-GURU with attention, and fully connected neural networks to perform early malware detection based on API calls predictions. As indicated in the results, we have set the initial length of the API call sequence to 20 and this sequence was fed to the fine-tuned model to predict the next 10, 20, and 30 API calls. As a result, we obtained 3 lengths of API call sequence (30, 40, and 50) which were fed to the trained model for testing. However, it is important to note the proposed fine-tuned model can also predict sequences longer than 30 API calls. 

Table \ref{result1} presents the results achieved by the proposed approach based on dataset 1. The proposed approach achieved an accuracy of 91.14\% and an AUC-ROC score of 0.9413 when detecting malware with an API call sequence length of 30. When the sequence length was increased to 40, the accuracy improved from 91.14\% to 95.39\% with an AUC-ROC score also increased to 0.9427. Finally, when the sequence length was increased to 50, the detection accuracy improved again to 95.97, resulting in an AUC-ROC score of 0.9477. These detection scores in Table \ref{result1} indicate that the proposed approach is accurate in predicting and detecting malware based on the behaviors of API calls. They also show that longer sequences of API calls generally lead to better performance. 

\begin{table*}[]
\centering
\caption{Classification accuracy and AUC-ROC score achieved by EarlyMalDetect on unseen samples from dataset 2.}
\label{result4}
\scalebox{0.86}{
\begin{tabular}{p{50mm}p{50mm}p{35mm}p{22mm}p{22mm}}

%\begin{tabular}{lllll}
\hline
Length of Initial Sequence &  Next Predicted API Calls & Total No.of API Calls & Accuracy (\%) & AUC-ROC Score \\ \hline
20                          & 10                                  & 30                                    & 82.86  &  0.8491 \\ \hline

20                          & 20                                  & 40                                    & 91.59     & 0.8736    \\ \hline

20                          & 40                                  & 50                                    & 92.39     & 0.9000   \\ \hline

\multicolumn{2}{l}{Average}                                      &  40                                    &  88.95   & 0.8742   \\ \hline
\end{tabular}
}
\end{table*}

\begin{table*}[!ht]
\centering
\caption{The precision, recall, and F1-score achieved by EarlyMalDetect on various lengths of API call sequences from the dataset 2.}
\label{result5}
%\resizebox{\columnwidth}{!}{%
%\begin{tabular}{|l|l|ll|ll|ll|}
\scalebox{0.75}{
%\begin{tabular}{|p{28mm}|p{55mm}|p{47mm}|}
\begin{tabular}{p{38mm}p{35mm}p{30mm}p{23mm}p{20mm}p{20mm}p{20mm}p{20mm}p{20mm}}
\hline
Length of Initial Sequence & Next Predicted API Calls & Total No.of API calls & \multicolumn{2}{c}{Precision}  & \multicolumn{2}{c}{Recall}  & \multicolumn{2}{c}{F1-Score} \\ \cline{4-9} 
&  &  & \multicolumn{1}{c}{Benign} & Malware  & \multicolumn{1}{c}{Benign} & Malware & \multicolumn{1}{c}{Benign} & Malware \\ \hline

20  & 10      &     30   & \multicolumn{1}{c}{} 0.9852 &  0.3278  & \multicolumn{1}{c}{} 0.8241   &  0.8741  & \multicolumn{1}{c}{}  0.8975 &  0.4768   \\ \hline 

20   & 20    &     40    & \multicolumn{1}{c}{} 0.9815    &  0.5187         & \multicolumn{1}{c}{} 0.9251      & 0.8222          & \multicolumn{1}{c}{} 0.9525         & 0.6361   \\ \hline 

20  & 30  &  50  &\multicolumn{1}{c}{} 0.9861 &  0.5467 & \multicolumn{1}{c}{} 0.9295    &  0.8667  & \multicolumn{1}{c}{} 0.9570  &  0.6705   \\ \hline

\multicolumn{2}{l}{Average} & 40 & \multicolumn{1}{c}{} 0.9842   & 0.4644 & \multicolumn{1}{c}{} 0.8929   &   0.8543 & \multicolumn{1}{c}{} 0.9356 &  0.5944 \\ \hline

\end{tabular}%
}
\end{table*}

\begin{table*}[h!]
\centering
\caption{Macro average and the weighted average for precision, recall, and F1-score achieved by EarlyMalDetect on dataset 2.}
\label{result6}
\scalebox{0.68}{
\begin{tabular}{p{38mm}p{35mm}p{30mm}p{20mm}p{18mm}p{18 mm}p{18mm}p{18mm}p{18mm}}
\hline
Length of Initial Sequence & Next Predicted API Calls & Total No.of API Calls & Macro P & Macro R & Macro F1 & Weighted P & Weighted R & Weighted  F1 \\ \hline

20 & 10  &  30   &  0.6565  &  0.8491   & 0.6871  & 0.9265   & 0.8286   & 0.8599   \\ \hline

20 & 20 &   40   & 0.7501  &  0.8737  &  0.7943  &   0.9401   & 0.9159 &   0.9242  \\ \hline

20 & 30 &  50    &  0.7664   & 0.8981   &   0.8137  &   0.9469   &  0.9239   &  0.9314  \\ \hline

\multicolumn{2}{l}{Average}    & 40 &  0.7243 &  0.8736   & 0.7650   & 0.9378 & 0.8894  & 0.9051  \\ \hline
\end{tabular}%
}
\end{table*}

\begin{figure*}[ht!]
  \centering
  \label{other-metrics-dataset1}
  \includegraphics[width=0.99\linewidth]{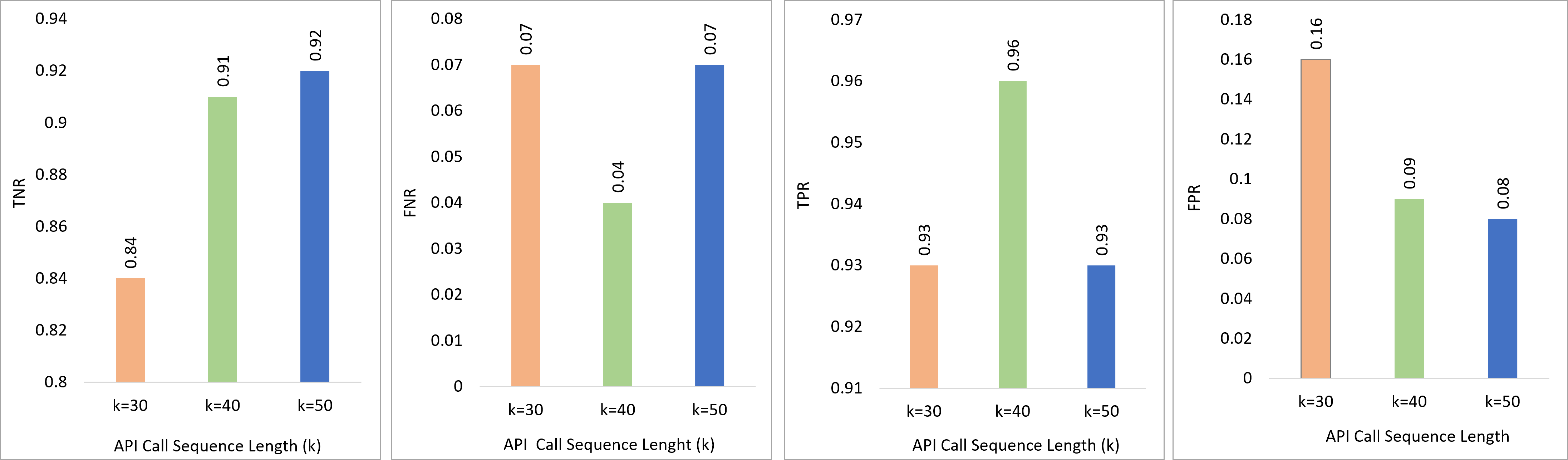}
  \caption{TNR, FNR, TPR, and FPR achieved by the proposed early malware detection approach on dataset 1.}
\end{figure*}

\begin{figure*}[!]
  \centering
  \label{other-metrics-dataset2}
  \includegraphics[width=0.99\linewidth]{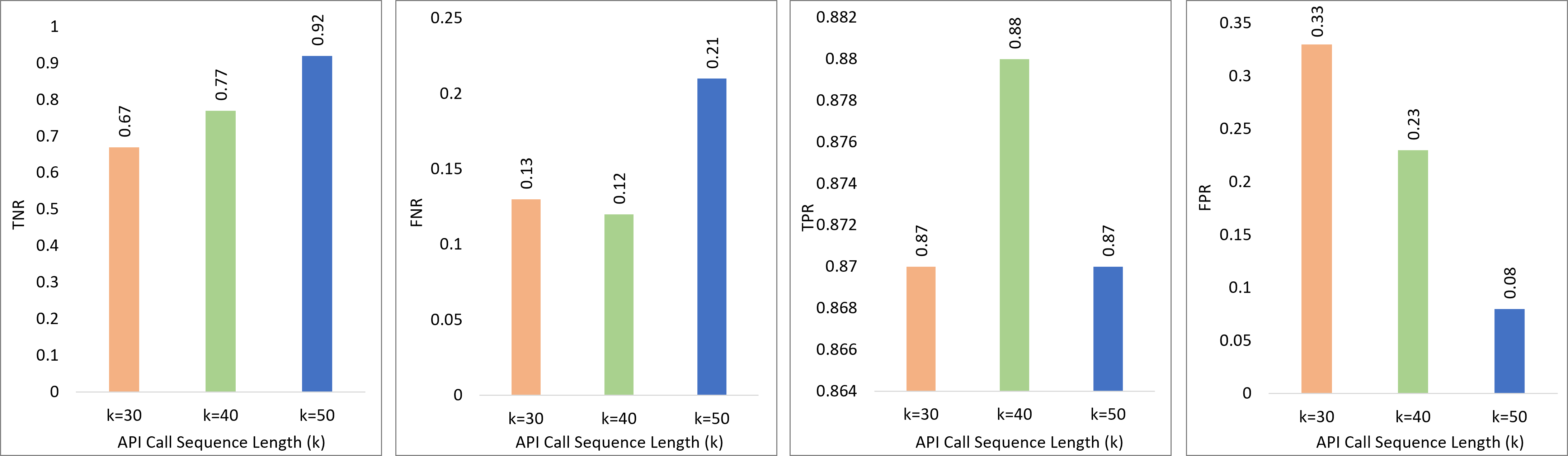}
  \caption{TNR, FNR, TPR, and FPR achieved by the proposed early malware detection approach on dataset 2.}
\end{figure*}

\begin{comment}
\begin{figure*}[ht!]
  \centering
  \label{other-metrics-alldataset}
  \includegraphics[width=0.99\linewidth]{accuracy-weighted F1-variation.png}
  \caption{Variation of accuracy and the weighted average for F1-score using (a and b) the dataset 1 and (c and d) using dataset 2.}
\end{figure*}

\end{comment}

The results in Table \ref{result2} show that with API calls sequence length of 30, the proposed approach (model) achieved a precision score of 0.9976 for benign and 0.5019 for malware.  The proposed approach also produced a recall score of 0.9049 for benign and 0.9778 for malware while it achieved an F1-Score of 0.9490 (for benign) and 0.6633 (for malware) with the same sequence length. Precision scores of 0.9932 for benign and 0.6562 for malware were achieved with a length of 40. The model also reached a recall score of 0.9521 for benign and 0.9333 for malware malware With the same sequence length (40). In addition, the model achieved an F1-Score of 0.9722 for benign and 0.7706 for malware.

On the other hand, the model obtained a precision score of 0.9933 for benign and 0.7079 for malware with a recall score of 0.9622 for benign and 0.9333 for malware on a sequence length of 50. In addition, an F1-Score of 0.9775 (benign) and 0.8051 (malware) was also achieved using the API call sequence of length 50. Overall, it can be observed that the precision scores for benign API calls are higher across all sequence lengths implying that the model is performing well in identifying benign API calls with a low rate of false positives. The results also indicate that the model correctly identifies malware samples, despite a few benign samples misclassified as malware.

Table \ref{result3} presents the obtained macro average and weighted average F1-score which shows that the model also performs well in both metrics (which is crucial for malware detection). This shows that the proposed model is more accurate in detecting malware while maintaining good performance. Furthermore, the weighted average for the F1-score increased from 0.9235 (using a sequence length of 30) to 0.9621 (with an API call sequence length of 50). The macro average recall remained consistent with the results indicating that the proposed model can identify most malicious activities. Furthermore, the  TNR, FNR, TPR, and FPR scores presented in Figures \ref{other-metrics-dataset1} and \ref{other-metrics-dataset2} show good performance achieved EarlyMaldetect (proposed model). In general, the results suggest that increasing the API call sequence length improves the performance of the malware detection model.

The proposed model has been also tested on dataset 2 and the results are presented in Tables \ref{result4} \ref{result5} and \ref{result6}. Upon examining the results in these tables, there is a slight drop in the performance compared to the results obtained from dataset 1. For instance, the proposed model achieved an accuracy of 92.39\% when detecting malware using dataset 2. The reduction in performance can be attributed to the concept drift, which refers to the changes in data distribution over time between the two datasets. That is, features and distributions of the data in the malware landscape can change significantly over time, i.e., the nature of malware changes as attackers develop new techniques and strategies to evade detection. Nevertheless, the overall results on both datasets prove significant performance achieved by the proposed model.

%This performance decrease can also be seen in Figure \ref{other-metrics-alldataset} which depicts the variation of accuracy and weighted average F1-score achieved using dataset 1 (released in 2018) and dataset 2 (released in 2021). 

%The next Section \ref{comparison-performance} provides insights into ElarlyMalDetect's performance against other  DL-based malware detection approaches. 

\begin{table*}[ht!]
\centering
\caption{Accuracy, false positive rate, false negative rate, and AUC-ROC achieved by different detection approaches on dataset 1.}
\label{result-table7}
\scalebox{0.89}{
\begin{tabular}{p{40mm}p{40mm}p{40mm}p{30mm}p{30mm}}
\hline
Detection Approach  & Accuracy (\%) &  False Positive Rate & False Negative Rate  & AUC-ROC Score  \\ \hline
          DistillBert-LSTM    & 94.38  &   0.06    &  0.05     & 0.9457\\ \hline
          DistillBert-CNN     & 94.64  &   0.06    &  0.02     & 0.9605 \\ \hline
          DistilBERT-BiLSTM   & 94.64  &   0.05    &  0.09     &  0.9305\\ \hline
          DistillBert-GRU     & 94.84  &   0.03    &  0.23     &  0.8681 \\ \hline
          EarlyMalDetect      & 95.97  &    0.08   & 0.07      & 0.9477 \\ \hline
\end{tabular}%
}
\end{table*}

\begin{table*}[ht!]
\centering
\caption{Precision, recall, and F1-score achieved by different models on dataset 1.}
\label{result-table8} 
\scalebox{0.82}{ 
\begin{tabular} {p{40mm}p{45mm}p{45mm}p{45mm}p{20mm}}
\hline
 Detection Approach  & Predicted Class & Precision & Recall & F1-Score \\ \hline
 
\multirow{2}{*}{DistillBert-LSTM}& Benign  & 0.9946    & 0.9434  &  0.9683  \\ %\cline{2-5} 
                                 & Malware   & 0.6214   &  0.9481 &   0.7507  \\ \hline
                                
\multirow{2}{*}{DistillBert-CNN}  & Benign  & 0.9977       &  0.9434     &  0.9698        \\ %\cline{2-5}
                                  & Malware   & 0.6286     &  0.9778     &   0.7652        \\ \hline

\multirow{2}{*}{DistilBERT-BiLSTM}  & Benign      &  0.9909   &  0.9499     &     0.9700          \\ %\cline{2-5} 
                                    & Malware     &  0.6406   &  0.9111      &    0.7523          \\ \hline
                                
\multirow{2}{*}{DistillBert-GRU} & Benign     &  0.9772         &  0.9659      &  0.9715     \\ %\cline{2-5}
                                 & Malware    &  0.6887         &0.7704        &  0.7273     \\ \hline
                                 
\multirow{2}{*}{EarlyMalDetect} & Benign &   0.9933      &   0.9622     &  0.9775   \\ %\cline{2-5}
                                                      & Malware    &    0.7079      &   0.9333    &  0.8051    \\ \hline
\end{tabular}%
}
\end{table*}

\begin{table*}[ht!]
\centering
\caption{Macro and the weighted average for precision, recall, and F1-score achieved by different models on Dataset 1.}
\label{result-table9}
\scalebox{0.72}{
\begin{tabular}{p{33mm}p{33mm}p{33mm}p{33mm}p{33mm}p{33mm}p{18mm}}
\hline
Detection Approach  & Macro P & Macro R & Macro F1 & Weighted P& Weighted R & Weighted F1\\ \hline
 DistillBert-LSTM & 0.8080   &    0.9458   &  0.8595   & 0.9613   &  0.9438  &  0.9489                  \\ %\hline
 DistillBert-CNN  &  0.8131  &   0.9606   &  0.8675    &    0.9647   &  0.9464  & 0.9515  \\ %\hline
 DistilBERT-BiLSTM &  0.8158   &  0.9305   &  0.8611 & 0.9596   & 0.9464    & 0.9505 \\ %\hline
 DistillBert-GRU  & 0.8330  & 0.8681 & 0.8494 & 0.9515  &  0.9484 & 0.9497 \\ %\hline
 EarlyMalDetect & 0.8506 &  0.9478 &  0.8913 & 0.9678 & 0.9597 & 0.9621 \\ \hline
\end{tabular}%
}
\end{table*}

\begin{table*}[h!]
\centering
\caption{Accuracy, false positive rate, false negative rate, and AUC-ROC achieved by different models on Dataset 2.}
\label{result-table10}
\scalebox{0.85}{
\begin{tabular}{p{40mm}p{40mm}p{40mm}p{30mm}p{26mm}}
\hline
Detection Approach & Accuracy (\%) &  False Positive Rate  & False Negative Rate  & AUC-ROC Score \\ \hline
           DistillBert-LSTM    & 91.39  &  0.07  &   0.21  &  0.8558  \\ %\hline
           DistillBert-CNN     &  90.60 &  0.09  &   0.14 &  0.8849 \\ %\hline
           DistilBERT-BiLSTM   & 91.86  &  0.07 &    0.15  & 0.8884 \\ %\hline
           DistillBert-GRU     & 91.00  &  0.09 &   0.13 & 0.8893 \\ %\hline
           EarlyMalDetect      & 92.39 &    0.08 &   0.21 & 0.9000 \\ \hline
           
\end{tabular}%
}
\end{table*}

\begin{table*}[ht!]
\centering
\caption{Precision, recall, and F1-score obtained when testing different models on Dataset 2.}
\label{result-table11} 
\scalebox{0.78}{ 
\begin{tabular} {p{40mm}p{45mm}p{45mm}p{45mm}p{16mm}}
\hline
Detection Approach  & Predicted Class & Precision & Recall & F1-Score \\ \hline
 
\multirow{2}{*}{DistilBERT-LSTM}  & Benign  &0.9778      &  0.9266    &   0.9515  \\ %\cline{2-5} 
                                  & Malware  &0.5121       &   0.7852    &     0.6199   \\ \hline

\multirow{2}{*}{DistillBert-LSTM} & Benign  &  0.9845  &  0.9251   & 0.9539        \\ %\cline{2-5}
                                  & Malware & 0.5275  &  0.8519   &   0.6516                  \\ \hline   

\multirow{2}{*}{  DistillBert-CNN} & Benign    &  0.9851  & 0.9106     &  0.9464        \\ %\cline{2-5} 
                                   & Malware   &   0.4854   &  0.8593   &   0.6203          \\ \hline
     
\multirow{2}{*}{DistillBert-BiLSTM} & Benign  &  0.9845  &  0.9251   & 0.9539        \\ %\cline{2-5}
                                    & Malware & 0.5275 &  0.8519   &   0.6516                  \\ \hline   

\multirow{2}{*}{DistillBert-GRU}   & Benign  &   0.9859         & 0.9121       & 0.9475  \\ %\cline{2-5} 
                                   & Malware  &   0.4916        & 0.8667       &  0.6273  \\ \hline

\multirow{2}{*}{EarlyMalDetect}   & Benign  &    0.9861        &  0.9295       & 0.9570  \\ %\cline{2-5} 
                                                & Malware  &   0.5467        & 0.8667       & 0.6705  \\ \hline
\end{tabular}%
}
\end{table*}

\begin{table*}[ht!]
\centering
\caption{Macro and the weighted average for precision, recall, and F1-score achieved by different detection approaches on dataset 2.}
\label{result-table12}
\scalebox{0.68}{
\begin{tabular}{p{33mm}p{33mm}p{33mm}p{33mm}p{33mm}p{33mm}p{14mm}}
\hline
Detection Approach  & Macro P & Macro R & Macro F1 & Weighted Pr & Weighted R & Weighted F1\\ \hline

DistilBERT-LSTM &  0.7449  & 0.8559 & 0.7857 & 0.9362 & 0.9140 &  0.9219  \\ %\hline

DistillBert-CNN  &  0.7352   &  0.8849 &  0.7833 & 0.9404 & 0.9060 & 0.9172  \\ %\hline

DistillBert-BiLSTM &  0.7560   &   0.8885  &   0.8027  & 0.9437  &   0.9186  &  0.9269 \\ %\hline

DistillBert-GRU  & 0.7387  &  0.8894   &  0.7874  & 0.9417 &  0.9080 &  0.9189 \\ %\hline

EarlyMalDetect  & 0.7664  &  0.8981  & 0.8137 &  0.9469  & 0.9239  & 0.9314 \\ \hline
\end{tabular}%
}
\end{table*}

\section{Performance Against Different Techniques}
\label{comparison-performance}
we have also implemented and evaluated the performance of different malware detection approaches against EarlyMalDetect. Each detection technique is tested using the same dataset to ensure consistency and fairness in the evaluation process. This is performed by analyzing and comparing various detection scores based on different metrics. By examining these detection models, we gain a better understanding of which technique may be the most suitable for detecting and preventing potential malware. Consequently, comparative results are presented in Tables \ref{result-table7}, \ref{result-table8}, \ref{result-table9}, \ref{result-table10}, \ref{result-table11}, and \ref{result-table12}. The detection approaches presented in this section are based on different deep learning architectures such as a combination of DistilBERT with other architectures like LSTM (Long Short-Term Memory), CNN (Convolutional Neural Network), GRU (Gated Recurrent Unit), and their bidirectional variants. 

The classification report in Table \ref{result-table7} shows that the DistillBert-LSTM detection approach(model) exhibits a balance between false positives and false negatives while maintaining high accuracy and a good AUC-ROC score, indicating a strong predictive capability. The DistillBert-CNN detection model has slightly higher accuracy than the DistillBert-LSTM. It also has a low FNR which means that it has fewer missed samples of the true positive class. On the other hand, the DistilBERT-BiLSTM model achieved the same accuracy as DistillBert-CNN, and it has a lower FPR. However, this model suffers from a relatively higher FNR, indicating more instances where it fails to recognize the positive instances. The AUC-ROC score is also lower compared to the first two models. Although the DistillBert-GRU detection model has the highest accuracy and lowest FPR (compared to the first three models), it has a considerably higher FNR, which means that it misses a substantial number of positive classifications. The AUC-ROC score is notably the lowest, which might indicate less consistency in the model's performance across different thresholds.  On the other hand, the proposed EarlyMalDetect model (DistillBert-BiGRU-Attention) shows the highest accuracy and a satisfactory AUC-ROC score, implying that it generally performs well over other models. However, it is important to mention that it has a higher FPR compared to other models.  In general, if we prioritize accuracy and a good balance between FPR and FNR, EarlyMalDetect is the most effective model. 

Table \ref{result-table8} presents the detection scores for precision, recall, and F1-Score. The results show that the DistillBert-LSTM model has a higher precision for benign instances but a lower precision for Malware instances. The F1 score balances precision and recall and is quite high for both classes. The CNN-based model improves upon the LSTM-based model slightly in precision for benign instances and significantly for the malware class, indicating fewer false positives when predicting malware. The recall for malware is very high and the model has a higher F1-score for the malware than the LSTM-based model. With the DistilBERT-BiLSTM model, there is a slight decrease in precision for the benign compared to the previous models, but there is an increase in precision score for malware samples. The recall is high, although it is slightly lower for malware instances compared to the CNN-based model. The DistillBert-GRU model has a decrease in precision for benign compared to other models, but it has an improvement in precision for malware and its recall is more balanced between the two classes. However, the F1-score for the malware class is lower among the models, suggesting some trade-offs in precision and recall. The EarlyMalDetect model appears to deliver the most balanced performance across precision, recall, and F1-scores, especially for the malware, which is typically the most challenging class to predict in anti-malware detection techniques. 

We have also evaluated the above models by measuring both macro and weighted averages. The obtained results are presented in Table \ref{result-table9}. The LSTM-based model has a good macro recall, but its macro precision is lower, indicating a higher number of false positives. The CNN-based model has slightly better macro precision and recall than the DistillBert-LSTM model. It appears to be strong at classifying instances correctly across different classes and it also performs well according to the obtained weighted metrics. The DistillBERT-BiLSTM model improves upon the macro precision, but it has a lower macro recall compared to the DistillBert-LSTM and DistillBert-CNN models. The DistillBert-GRU model achieves the highest macro precision, but it has the lowest macro recall among the models. Its weighted precision is lower, but it maintains a high weighted recall and F1-score. The EarlyMalDetect model outperforms all other models in both macro and weighted averages across all three metrics. It is also important to highlight that the classification reports presented in Tables \ref{result-table10}, \ref{result-table11}, and \ref{result-table12}, also prove that the proposed model performs well over other detection models.

\subsection{Limitations of this Work}
\label{ourmodel-limitations}
Although this work has fine-tuned the proposed technique using 69,234 sequences representing malware and benign executable programs, the dataset size is still small given the large size of malware programs which is rapidly increasing. Thus, in future work, we will improve the performance of the proposed approach by increasing the size of the training set and then refine-tune the proposed model on large samples of API Call sequences. While the fine-tuned model can automatically generate new sequences and predict the next API calls in a given sequence, it is important to note that the fine-tuned model can generate irrelevant sequences in some cases. This is because language-based models can be biased when performing certain prediction tasks \cite{tamkin2021understanding}. It is worth mentioning that, we will also plan to test the proposed approach on other large transformer models such as GPT-3.5 or GPT-4 on malware detection in Windows. 
 
\section{Conclusion}
\label{concl-future}  
In this work, we have presented a novel preventive and mitigation approach for early detection of malware attacks based on API call sequence prediction. Our proposed fine-tuned transformed model has demonstrated significant potential in accurately predicting the behaviors of malicious programs in Windows. The proposed approach is mainly based on sequence prediction with transformer models, bidirectional GRU with attention, and fully connected neural networks. Our results demonstrate better performance in detecting and classifying unknown Windows malware, making the proposed approach highly effective compared to the existing malware detection approaches. We performed a comparison between different detection models and the proposed approach demonstrated superior performance against them. By providing a preventive approach for predicting and detecting malicious behaviors of malware programs before they can infect the victim's system, the proposed approach can help to improve malware detection. The results provide new insights into malware detection in Windows. Therefore, we believe this work will be valuable to the scientific community and potentially impact the real-world application of malware detection in Windows.   

\bibliographystyle{IEEEtran}
%\bibliographystyle{plainnat} % use this to have URLs listed in References
%\cleardoublepage
\bibliography{IEEEabrv,references}

%\begin{comment}

\section*{BIOGRAPHY}
%\label{biography}
%\vfill
\vspace{11pt}
\textbf{Pascal Maniriho} received his B.Tech with Honors in Information and Communication Technology from Umutara Polytechnic, Rwanda, and Master’s degree in Computer Science from Institut Teknologi Sepuluh Nopember (ITS), Indonesia, in 2013 and 2018, respectively.  He has been working in academia in  Information Technology since 2019. He is currently pursuing his Ph.D. degree in cybersecurity at La Trobe University, Australia. His research interests include malware detection, data theft prevention, information security, machine learning, and deep learning.

\textbf{Abdun Naser Mahmood} received the B.Sc. degree in applied physics and electronics, and the M.Sc. (research) degree in computer science from the University of Dhaka, Bangladesh, in 1997 and 1999, respectively, and the Ph.D. degree from the University of Melbourne, Australia, in 2008. He is currently an Associate Professor with the Department of Computer Science, School of Engineering and Mathematical Sciences, La Trobe University. His research interests include data mining techniques for scalable network traffic analysis, anomaly detection, and industrial SCADA security. He is a senior member of the IEEE.

\textbf{Mohammad Jabed Morshed Chowdhury} is currently working as an Associate Lecturer at La Trobe University, Melbourne, Australia. He has earned his Ph.D. Candidate at Swinburne University of Technology, Melbourne, Australia. He has earned his double Masters in Information Security and Mobile Computing from Norwegian University of Science and Technology, Norway, and the University of Tartu, Estonia under the European Union’s Erasmus Mundus Scholarship Program. He has published his research in top venues including TrustComm, HICSS, and REFSQ. He is currently working with Security, Privacy, and Trust. He has published research work related to blockchain and cyber security in different top venues.

\vfill
    
%\end{comment}

\end{document}